# One-step synthesis of mesoporous Cobalt sulfides ($CoS_x$) on the metal substrate as an efficient bifunctional electrode for overall water splitting

Haidong Bian [a,b], Tongyuan Chen [a], Zhixuan Chen [a], Zebiao Li [c,d], Peng Du [c,d], Binbin Zhou [c,d], Xierong Zeng [a], Jiaoning Tang [a], Chen Liu [a,*]

[a] Shenzhen Key Laboratory of Polymer Science and Technology, College of Materials Science and Engineering, Shenzhen University, Shenzhen 518055, PR China

[b] Key Laboratory of Optoelectronic Devices and Systems of Ministry of Education and Guangdong Province, College of Optoelectronic Engineering, Shenzhen University, Shenzhen 518055, PR China

[c] Hong Kong Branch of National Precious Metals Material Engineering Research Centre, City University of Hong Kong, Hong Kong, China

[d] Department of Materials Science and Engineering, City University of Hong Kong, Tat Chee Avenue 83, Kowloon, Hong Kong, China

* E-mail: liuchen@szu.edu.cn

**Abstract**

Electrocatalysts based on transition metal sulfides (TMSs) are drawing accelerating concerns in renewable energy research because of their intrinsically excellent activities towards both hydrogen evolution reaction (HER) and oxygen evolution reaction (OER). To date, considerable efforts are made to improve the performance of these catalysts, but ignoring the improper synthesis strategy would incur additional cost to the catalyst. Herein, a convenient, one-step anodization method is developed for fast construction of cobalt sulfides ($CoS_x$). Without any high-temperature or long-time treatment, mesoporous $CoS_x$ is self-grown on the metal substrate in minutes. As a result, as-anodic $CoS_x$ requires overpotentials of 102 mV for HER and 284 mV for OER to achieve a current density of 10 mA m$^{-2}$ in alkaline solution. Moreover, the tandem bifunctional as-anodic $CoS_x$ exhibits a required cell voltage of 1.64 V for overall water splitting in alkaline solution, exceeding most of the documented Co-based electrocatalysts.

## 1. Introduction

Transition metal sulfides (TMS, where M = Fe, W, Co, V, Mo or Ni) have attracted growing global concerns towards electrochemical water splitting (both oxygen evolution reaction (OER) and hydrogen evolution reaction (HER)), owing to their low cost, abundant resources and fascinating electrocatalytic performance, making them promising alternatives of precious metal and metal oxide catalysts (Pt, Pd or their alloys for HER catalysis, and $RuO_2$, $IrO_2$ for OER catalysis).[1-4] Amongst, cobalt sulfides, including CoS,[5] $Co_3S_4$,[6] $CoS_2$,[7] $Co_9S_8$,[4] $Co_xS_y$,[8] etc. arouse tremendous interest due to their low price and low reaction barrier for HER and OER. To achieve the intrinsic high activity and efficiency of the catalyst, cobalt sulfides with various morphologies and structures have been prepared and investigated.[9-12] However, most of the present works pay much attention to pursuing the high performance of the catalysts, while ignoring their fabrication methods. For fundamental research in the lab, it is accessible, but for the industrial application we have to realize the cost of the catalyst should include the preparation cost of the catalysts, as an improper synthesis procedure may incur extra tremendous cost to the catalyst.

To date, the most common fabrication strategy is through hydrothermal/solvothermal[6, 13-15] or high-temperature sulfuration process.[5, 16] Using these methods, high temperature and long-time reactions are needed, which inevitably consume huge external energy and increase the manufacturing costs of the catalysts. Other methods, such as electrospinning,[17] anion exchange reaction,[18] hot injection,[19] plasma-induced dry exfoliation,[11] template-assisted[20] and microwave-assisted[21] method, have also been developed for different cobalt sulfide nanostructures. These methods usually entail either tedious fabrication processes or expensive facilities. In addition, the obtained catalysts are generally in the form of a powder and need to be pasted on the substrate for electrochemical tests, which would further increase the cost of the electrode and the difficulty of recycling the catalysts. Actually, in many of these cases, when considering the fabrication cost, the price of the catalyst is even higher than precious metal and metal oxides. Thus, to develop a facile, cheap method for constructing cobalt sulfide, while maintaining its electrocatalytic activity, is still of great importance and needs more efforts.

Electrochemical deposition can provide a cheap and quick process to construct cobalt sulfides on the substrate.[22] However, the weak combination between the active materials and substrate would deteriorate the long-term reaction of the catalysts. Anodization, as another important electrochemical method, has provided an efficient strategy for construction various robust nanostructures on the substrate. However, to date, only cobalt oxides are directly obtained after anodization process.[23-25] Herein, for the first time, a mesoporous cobalt sulfide ($CoS_x$) structure was fabricated via this facile, one-step anodization method, with merely treating the Co foil in the electrolyte for 5 min. This porous, robust, self-grown nanostructure not only brings high surface active sites for catalytic reaction, but also facilitates electron transport across the active material. As a result, the as-anodic $CoS_x$ catalyst exhibits both outstanding HER and OER performance in 1.0 M KOH electrolyte. The as-anodic $CoS_x$ also demonstrates a superior overall water splitting performance, comparable to Pt||RuO$_2$ couple electrodes in alkaline electrolyte. Moreover, deactivated CoS$_x$ after long-time usage can be easily removed and the residual metal substrate can be anodized, making one piece of the Co foil being used many times. Given the similarly shared mechanism among the various TMS obtained from their metal substrates, the anodization strategy developed in this work opens up a quick, cheap and efficient way to construct transition metal sulfides (TMS) as efficient electrocatalysts towards water splitting.

## 2. Experimental

*2.1 Anodic fabrication of nanoporous CoS$_x$*

CoS$_x$ nanoporous structure was directly fabricated on a cobalt foil via a facile anodization method described in our previous work.[26-28] Prior to anodization, the cobalt sheet was ultrasonically cleaned in acetone, distilled water and ethanol, and then dried in air. The experiment was carried out in a home-made two-electrode cell with cobalt sheets (> 99 wt%, Fuxin Metal Products LTD, Baoji, 0.5 * 1 cm$^2$ as the working area) as the anode and Pt coil as the counter electrode. The electrolyte was an ethylene glycol (EG) (≥99.8 wt%, International Laboratory USA) solution with addition of 0.05 M sodium sulfide (Na$_2$S, ≥96.0 wt%, International Laboratory USA). A constant voltage of 10 V was applied for 5 min by a digital regulated DC power supply (eTM-L305SPL). After anodization, the samples were taken out, carefully rinsed with ethanol and then dried in vacuum at 40 °C. Thermal annealing process was carried out at 400 °C for 1 h in Ar with a heating rate of 10 °C/min.

*2.2 Material characterization*

X-ray diffraction (XRD) patterns were recorded on an X-ray diffractometer (XRD, Rigaku) with Cu-K$_\alpha$ radiation (λ= 1.5405 Å). Surface morphology observation and chemical composition were carried out on a field-emission scanning electron microscope (FEI APREO S(A5-112)) with an energy dispersive X-ray

(EDX) detector (Oxford INCA). Transmission electron microscope (TEM) and high-resolution transmission microscope (HRTEM) images were observed on a JEM-200 (JEOL) electron microscopy equipped with an EDX detector (Oxford INCA). The surface chemical states and composition of the obtained samples were analyzed by an X-ray photoelectron spectroscope (XPS, VG ESCALAB 220i-XL) with all the peak positions calibrated with respect to the C 1s peak at 284.8 eV. The surface area of the samples were tested on a Quantachrome Nova 1200e Surface Area Analyzer and evaluated by Brunauer-Emmett-Teller method (BET).

*2.3 Electrochemical measurements*

All the electrochemical activities were performed on a Princeton electrochemical workstation (VERSASTAT 3) in a conventional three electrode setup, with the as-fabricated $CoS_x$ electrode (with exposing area of 0.5 cm × 1.0 cm and loading weight of ~ 1.1 mg cm$^{-2}$) as working electrode, a saturated Ag/AgCl electrode as reference electrode, and a commercial Pt foil (1 cm$^2$) as counter electrode. The bare Co foil, commercial Pt foil were also tested for HER as a comparable result. For OER, bare Co foil and commercial $RuO_2$ catalysts were prepared for comparison. The commercial $RuO_2$ catalytic ink was prepared by dispersing active material (5 mg) in a 1 mL mixed isopropanol/water (3:1 $v/v$) solvent and 30 μL 5 wt % Nafion solution. Then, the $RuO_2$ catalytic ink was deposited onto the Co foil with a catalyst loading of ~ 1.0 mg cm$^{-2}$. All the electrochemical tests towards the HER and OER were evaluated in 1 M KOH (pH = 13.8). Electrochemical impedance spectroscopy (EIS) measurements were carried out with an amplitude of +/-5 mV in the frequency range of 100 kHz - 0.01 Hz at the desired overpotential ($\eta$) of both HER and OER. A linear sweep voltammetry (LSV) with a scan rate of 2 mV s$^{-1}$ was applied to evaluate the activity of the obtained materials. Prior to LSV testing, the iR loss compensation was corrected, where i is the test current and R is the compensation resistance. Cyclic voltammetry (CV), chronoamperometric and chronopotentiometric tests were carried out to examine the stability of the working electrode. All potentials

were converted to reversible hydrogen electrode (RHE) according to the equation:[29]

$$E \text{ (vs RHE)} = E \text{ (vs Ag/AgCl)} + 0.197 + 0.059 \times \text{pH} (=13.8)$$

where E (*vs* Ag/AgCl) was the recorded potential *vs.* reference electrode.

## 3. Results and discussion

The self-grown process of cobalt sulfides ($CoS_x$) on the substrate was followed a oxidation-deposition process, as our previous report.[28] When the electric field was exerted on the Co foil, Co ions ($Co^{2+}/Co^{3+}$) released from the metal surface reacted with the transported sulfide ions ($S^{2-}$) to deposit $CoS_x$ on the surface (as illustrated in Fig. 1a). It should be mentioned that the anodization environment and anodization parameters should be carefully controlled in case of forming cobalt oxides. In this experiment, $CoS_x$ was successfully fabricated at 10 V for 5 min in $Na_2S$-contained anhydrous ethylene glycol (EG) solution. A porous layer comprised of wrinkled $CoS_x$ nanostructures was observed on the surface for the as-anodic $CoS_x$, as shown in Fig. 1b. After high-temperature treatment (400 °C 1 h in Ar), the porous structure was retained but slightly changed with a more condensed surface (Fig. S1a-b). The XRD patterns were collected for the as-anodic and Ar-annealed samples (Fig. 1c and d), as the peaks at 41.7°, 44.7°, 47.5°, 62.7° and 75.9° were ascribed to (100), (002), (101), (102) and (110) planes of Co substrate (JCPDS card No. 05-0727). A slight broad peak at ~ 30° indicates the amorphous feature of as-anodic sample, while peaks at 29.8°, 31.1°, 39.5°, 51.9°, 61.8° and 50.2°, 62.2° observed for Ar-annealed sample were attributed to (311), (222), (331), (440), (622) planes of $Co_9S_8$ (JCPDS card No. 02-1459) and (511), (620) planes of $Co_3S_4$ (JCPDS card No. 42-1448), respectively. The morphology and structure of as-anodic and Ar-annealed $CoS_x$ were further investigated by TEM and HRTEM. A porous structure was clearly observed for the as-anodic and Ar-annealed $CoS_x$, as shown in Fig. 2a and d. Detailed observation of as-anodic $CoS_x$ demonstrated that some tiny pores (below 10 nm) were presence in the ligament of the macropores (as arrows indicated in Fig. 2b). The existence of these tiny pores would contribute a higher surface area and more surface atoms in the as-anodic $CoS_x$ than that of Ar-annealed $CoS_x$. No obvious lattice fringes was observed in the HRTEM image (Fig. 2c) of as-anodic CoS$_x$, indicating the amorphous feature of the as-fabricated sample, which was consistence with the XRD result. The corresponding TEM-EDX elemental mapping (Fig. S2) showed the element S and Co were homogeneously distributed throughout the porous structure, with the composition of S and Co confirmed to be 56.9 at.% and 43.1 at.%, respectively. In comparison, clearly crystal lattices were observed in the HRTEM image of Ar-annealed sample (Fig. 2f). The inter-plane distance of 0.30 nm was indexed to the (311) plane of Co$_9$S$_8$,[30] while the interlayer spacing of 0.285 and 0.165 nm could be assigned to the (311) and (440) planes of Co$_3$S$_4$,[6] respectively.

The surface chemical composition and bonding states of the as-anodic and Ar-annealed CoS$_x$ were investigated by XPS. The high-resolution Co 2p spectra of the samples were split into three groups of spin-orbit peaks, as shown in Fig. 3a. The fitted curves demonstrated the coexistence of Co(II) and Co(III) for both the as-anodic and Ar-annealed samples, and the binding energies at 796.4 eV and 780.3 eV were assigned to Co(III) 2p$_{1/2}$ and Co(III) 2p$_{3/2}$, and the binding energies at 797.9 eV and 782.2 eV correspond to the Co(II) 2p$_{1/2}$ and Co(II) 2p$_{3/2}$, respectively.[31] Meanwhile, the peaks at 802.5 eV and 786.2 eV were the satellite peaks of Co 2p$_{1/2}$ and 2p$_{3/2}$ spin orbits, respectively.[32] As for the S 2p core-level XPS spectrum (Fig. 3b), the split peaks at 162.6 eV and 161.4 eV were ascribed to the S 2p$_{1/2}$ and 2p$_{3/2}$ orbitals of Co-S, while the peak at 167.2 eV corresponded to the formation of oxidized S species on the surface.[33, 34] Furthermore, N$_2$ adsorption-desorption isotherms were obtained for as-anodic and Ar-annealed CoS$_x$ (Fig. 3c). Typical Type IV characteristics with distinct hysteresis loops were clearly displayed, suggesting the porous feature of both samples. The surface areas of as-anodic and Ar-annealed CoS$_x$ were estimated by the Brunauer-Emmett-Teller (BET) method, to be 35.6 and 15.6 m$^2$ g$^{-1}$, respectively. Fig. 3d showed the pore size distribution curve of as-anodic CoS$_x$ (obtained by the adsorption isotherm) by the Barrett-Joyner-Halenda (BJH) method, revealing three types of pores (< 5 nm, ~ 10 nm and ~ 20 nm) were mainly contained in the porous as-anodic CoS$_x$. However, for Ar-annealed CoS$_x$, the majority of the pores were around 10 nm (Fig. S3). The diminished pores could be ascribed to the structure merge and rearrangement during heat treatment process at high temperature. The difference of pore size and surface area resulted in different active sites and atoms arrangement on the surface, which was related to the catalytic activity of the catalysts towards HER and OER.

The electrochemical HER performance of the as-anodic and Ar-annealed $CoS_x$ catalysts were evaluated via electrochemical testing with a standard three-electrode system at ambient temperature. The polarization curves of the samples were obtained by linear sweep voltammetry (LSV) (Fig. 4a). As expected, Pt showed the lowest onset overpotential (defined as $\eta_o$ at 1 mA cm$^{-2}$) of 2.8 mV, owing to the intrinsic superior performance of Pt towards HER. For as-anodic $CoS_x$, it exhibited an onset overpotential of 9.7 mV, much lower than that of Ar-annealed $CoS_x$ (16.4 mV), indicating an inferior performance after heat treatment. Fig. 4b summarized the overpotential ($\eta$) of all the samples at various current densities (10 ~ 50 mA cm$^{-2}$). The as-anodic $CoS_x$ catalyst exhibited a low $\eta$ of 102 mV at 10 mA cm$^{-2}$, while the Ar-annealed $CoS_x$/Co showed a much higher $\eta$ of 192 mV. By comparison, Pt foil and Co foil delivered a $\eta$ of 56 and 388 mV at 10 mA cm$^{-2}$, respectively. At a high current density of 50 mA cm$^{-2}$, the as-anodic $CoS_x$ exhibited the best electrocatalytic performance with a $\eta$ of 200 mV, outperforming the Pt foil (214 mV). Additionally, at high current densities (> 89 mA cm$^{-2}$), the Ar-annealed $CoS_x$ also outperformed the Pt catalyst. The not impressive performance of Pt at higher current densities was ascribed to the limited surface atoms of the bulk Pt foil.

The HER kinetics of the fabricated electrodes was further estimated by their Tafel plots (Fig. 4c), which indicated the change of $\eta$ when the current density was decreased or increased 10-fold. The Tafel plots were fitted according to the Tafel equation ($\eta$ = a + b log |$j$|, where $\eta$ was the overpotential, a was the constant, b was the Tafel slope and $j$ was the current density). Typically, the Tafel analysis of the as-anodic $CoS_x$, Ar-annealed $CoS_x$, Co and Pt catalysts delivered a Tafel slope of 92, 132, 251 and 59 mV dec$^{-1}$, respectively. The smaller slope value of as-anodic $CoS_x$ compared to Ar-annealed $CoS_x$ suggested a faster charge transfer for an increased HER velocity. Additionally, as an intrinsic characteristic of the electrocatalysts, the Tafel slope could also be used to predict the reaction mechanism of the HER process. Mechanistically, in alkaline medium, electrochemical HER process proceeded through three possible steps for the reduction of water molecules to $H_2$ on the surface of a catalyst.[35, 36] The initial step was the Volmer adsorption process, generating adsorbed hydrogen atom (H*) on the active material surface (M) through water molecule reacting with an electron ($H_2O + M + e^- \rightarrow M\text{-}H^* + OH^-$). The following steps had two different ways for $H_2$ generation: electrochemical desorption step (Heyrovsky reaction, $H_2O + e^- + M\text{-}H^* \rightarrow H_2 + OH^- + M$) and chemical desorption step (Tafel reaction, $2M\text{-}H^* \rightarrow H_2 + 2M$). As indicated by the value of Tafel slope (92 mV dec$^{-1}$), the Volmer step should be the rating-determining step for the HER process of as-anodic $CoS_x$ electrode.

The electron transfer kinetics were evaluated by EIS spectrum measured in the frequency range between 100 kHz and 0.01 Hz at an overpotential of 250 mV under HER conditions (Fig. 4d). The fitted equivalent circuit comprised of a charge-transfer resistance ($R_{ct}$), an internal resistance ($R_s$) and a constant phase element (CPE) (inset of Fig. 4e). $R_{ct}$ represented the charge transfer resistance at the electrode/electrolyte interface, while $R_s$ was the ohmic resistance arising from electrolyte and all contact. All the fitted parameters were summarized in Table S1. It was observed that all the electrodes exhibited a $R_s$ of 2 ~ 4 Ω, meaning an excellent electrical conductivity of all the electrodes. The similar $R_s$ of as-anodic and Ar-annealed $CoS_x$ also indicated the heat treatment would not great change the conductivity of amorphous $CoS_x$, which may be due to the intrinsic excellent conductivity of cobalt sulfides either in amorphous or in crystalline state. Moreover, the $R_{ct}$ of as-anodic $CoS_x$ (5.58 Ω) was much lower than that of Ar-annealed $CoS_x$ (7.94 Ω), Co (266 Ω), Pt (5.68 Ω) and most of cobalt sulfides obtained by other methods (such as CoS (65.2 Ω, hydrothermal),[37] $CoS_2$-ns/CFP (25.6 Ω, electrodeposition),[38] and a-CoS (262 Ω, hydrothermal),[39] indicating a much faster charge transport rate across the interface of the catalyst towards HER process.

As another essential factor in estimating the performance of electrocatalysts, electrochemical stability was evaluated by accelerated cyclic voltammetry (CV) cycling test (100 mV/s). A slight change of polarization curves after 2000 cycles (Fig. 4e) was observed, demonstrating the robustness of the electrode. The inset in Fig. 4e was the long-term chronoamperometry curve ($j$-t curve), which showed the current change with the time of the as-anodic $CoS_x$ catalyst at a constant overpotential of 50 mV. After 25 h's testing, no obvious current density decay was observed, suggesting the outstanding electrochemical stability of the electrode. The morphology and structure after long-time testing was depicted in Fig. S4 a-b, from which the originally porous structure and amorphous feature of as-anodic $CoS_x$ were preserved, indicating the stability of as-anodic $CoS_x$ towards long-term HER process. Additionally, chronopotentiometry with different current densities (10, 20, 50 and 100 mA cm$^{-2}$) was also used to analysis the stability of the catalyst. The voltage exhibited a sluggish change over long-time testing at different current densities, again demonstrating the durability of the catalyst (Fig. 4f). Table S2 compared the different fabrication methods for cobalt sulfide-based catalysts and their HER performance. The as-anodic $CoS_x$ exhibited an excellent performance towards HER, which was superior to most of reported cobalt sulfides-based catalysts. Moreover, as-anodic $CoS_x$ catalyst also exhibited its advantages in synthesis procedures, merely with exerting a potential of 10 V for 5 min at room temperature, while avoiding any high-temperature, long-time treatment or tedious fabrication steps.

The electrocatalytic activities towards OER for as-anodic and Ar-annealed $CoS_x$ were also assessed in a typical three-electrode setup in 1.0 M KOH. Different from HER process, metal chalcogenides usually went through oxidation reactions and phase transformation to oxide or oxyhydroxide on the surface during OER process,[40-42] which caused severe current change in the first or first several cycles of polarization curves. Thus, one had to pay much attention to the reported overpotential values when using these polarization curves. In this work, significant oxidation peaks were observed in the first cycle of CV curves of as-anodic

CoS$_x$ (Fig. S5a), and anodic peaks at ~ 1.07, 1.30 and 1.45 V were ascribed to the oxidation of Co (I) to Co (II), Co (II) to Co (III) and Co (III) to Co (IV), respectively.[43-45] No severe oxidation peaks shift and current density change were seen from the second cycle in the CV curves (Fig. S5b), indicating a much steadier state after the first cycle. Hence, to exclude or reduce the effect of faradic current from the oxidation reaction, the electrochemical characterization of all the electrodes towards OER process were recorded after several CV cycling testing at a scan rate of 100 mV s$^{-1}$.

Fig. 5a showed the OER LSV curves for as-anodic CoS$_x$, Ar-annealed CoS$_x$, bare Co and the commercial RuO$_2$. The as-anodic CoS$_x$ catalyst required a much lower $\eta$ (284 mV) than that of Ar-annealed CoS$_x$ (337 mV) and bare Co (476 mV) to achieve a current density of 10 mA cm$^{-2}$ (Fig. 5b). Additionally, the as-anodic CoS$_x$ catalyst displayed a remarkable catalytic activity ($\eta_{50}$ = 362 mV) at increased current densities (> 50 mA cm$^{-2}$), outperforming the OER performance of commercial RuO$_2$ ($\eta_{50}$ = 363 mV). The low overpotential value of as-anodic CoS$_x$ ($\eta_{10}$ = 284 mV) was still outstanding among the best reported cobalt sulfide-based catalysts (Table S3), including CoS$_2$/CC (291 mV),[46] CoS-RGO (350 mV),[5] CoS$_2$ (386 mV),[47] Co$_3$S$_4$ (363 mV),[18] and CoS (307 mV).[19] The OER kinetics of all the catalysts were estimated based on the Tafel slopes (Fig. 5c). The as-anodic CoS$_x$ showed a lower Tafel slope (75.8 mV dec$^{-1}$) than those of Ar-annealed CoS$_x$ (102.6 mV dec$^{-1}$) and Co (178.8 mV dec$^{-1}$), suggesting a more rapid OER kinetics of the as-anodic CoS$_x$. From the Tafel slope value, the first electron transfer step (M + OH$^-$ → MOH + e$^-$) was the rate-determining step for OER process of as-anodic CoS$_x$. Moreover, the EIS Nyquist plots of all the electrodes were conducted at a potential of 1.63 V (overpotential of 400 mV) vs. RHE (Fig. 5d). The fitted equivalent circuit (inset of Fig. 5d) demonstrated a decreased $R_{ct}$ value (1.06 Ω) for as-anodic CoS$_x$ compared to those of Ar-annealed CoS$_x$ (1.18 Ω), Co (6.8 Ω) and RuO$_2$ (2.68 Ω) (Table S1). The low $R_{ct}$ value indicated a rapid charge transfer in the as-anodic CoS$_x$ electrode, thereby delivering a high OER activity.

To better clarify the electrocatalyst's intrinsic activity, electrochemically active surface area (ECSA) is an important parameter, which could be evaluated from the double layer capacitance ($C_{dl}$). To achieve $C_{dl}$, CV measurements with different scan rates from 10 to 100 mV s$^{-1}$ were carried out between a potential window of 0.26 – 0.36 V *vs.* RHE (Fig. S6). The half current difference between cathodic and anodic process at 0.31 V were plotted vs. the scan rates (Fig. 5e). Then, $C_{dl}$ was calculated by the slope of the linear relationship between the current density against the scan rate. The $C_{dl}$ of as-anodic CoS$_x$ was 41.8 mF cm$^{-2}$, approximately 3 times to that of Ar-annealed CoS$_x$ (14.7 mF cm$^{-2}$). It was consistent with the previous BET surface area. As $C_{dl}$ was proportional to the ECSA of the catalysts, the $C_{dl}$ value also indicated that the heat-treatment process decreased the catalytic active sites of as-anodic CoS$_x$. Thus, as-anodic CoS$_x$ exhibited a superior electrocatalytic activity towards both HER and OER in comparison to Ar-annealed CoS$_x$.

The long-term durability of as-anodic CoS$_x$ electrode was evaluated by CV cycling tests at a scan rate of 100 mV s$^{-1}$. No obvious shift of polarization curves after 2000 cycles (Fig. S7a) was observed, demonstrating the robustness of the electrode. The chronoamperometric test with an overpotential of 50 mV showed a slight current decay after 25 h (Fig. S7b), also demonstrating the stability of the electrode. Furthermore, chronopotentiometric curve with different current densities ranged from 10 to 100 mA cm$^{-2}$ was also carried out to investigate the durability of as-anodic CoS$_x$ catalyst (Fig. 5f). After continuous testing (20 h), the potentials remained stable at different current densities. However, when the current density was back to 10 mA/cm$^2$ after a long period of 20 h, the potential slightly decreased and then stabilized at a lower value. The change of potential after long-time OER process could be ascribed to the formation of cobalt oxides or oxyhydroxides on the surface after long-time reaction, which has already been confirmed by some previous research.[40, 48] Briefly, the wrinkled morphology was retained after long-time OER process as confirmed by TEM observation (Fig. S8a-b). A lattice distance of 0.142 nm was clearly seen on the outer surface, which was demonstrated to be (110) plane of CoOOH.[49] This in-situ phase transformation from sulfides to oxyhydroxides, which actually acted as catalytic active sites, together with the formation of CoOOH/CoS$_x$ heterojunctions, had endowed the catalysts with slightly improved OER performance after long-time reaction.

Fig. 6a compared the electrocatalytic performance ($\eta_{10}$) of some latest reported cobalt-based compounds as bifunctional catalysts for overall water splitting.[5, 6, 9, 19, 37, 50-59] In the figure, these bifunctional catalysts were roughly separated into four different groups according to their sum overpotential value ($\eta_{10}^S$) of HER and OER ($\eta_{10}^S = \eta_{10}^{HER} + \eta_{10}^{OER}$). The catalysts with a $\eta_{10}^S$ value below 300 mV were in group I, indicating both fantastic HER and OER performance. In group II, the catalysts exhibited outstanding HER and OER performance, delivering a $\eta_{10}^S$ value between 300 mV and 400 mV. Most of the statistical catalysts were in group III with $\eta_{10}^S$ between 400 mV and 500 mV, suggesting the catalysts possessed either an excellent HER or OER activity. For the catalysts with a $\eta_{10}^S$ value higher than 500 mV (region IV), modification of the catalysts should be carried out for further overall water splitting application. Briefly, the catalyst from group I and II exhibited potential application as a bifunctional catalyst towards both OER and HER, while the catalysts in group III were only suitable for half electrode reaction either in HER or in OER.

Encouraged by the outstanding HER and OER performance of as-anodic CoS$_x$ catalyst, as well as its feasible synthesis method, the as-anodic CoS$_x$ electrode was employed as both the anode and cathode for overall water splitting testing. As a result, to achieve a current density of 10 mA m$^{-2}$ at a scan rate of 2 mV s$^{-1}$, the as-anodic CoS$_x$||as-anodic CoS$_x$ required a cell voltage of 1.64 V, which was much lower than that of Ar-annealed CoS$_x$|| Ar-annealed CoS$_x$ (1.76 V) and comparable to that of noble Pt||RuO$_2$ electrode (1.61 V) (Fig. 6b). Additionally, in comparison to most previous reported bifunctional Co-based electrocatalysts (Table S4), the as-anodic CoS$_x$ demonstrated its quick and efficient accessibility as well as its promising electrocatalytic performance. Furthermore, the constructed as-anodic CoS$_x$||as-anodic CoS$_x$ electrolyzer demonstrate a remarkable stability at a potential of 1.6 V (Fig. 6c). A remarkable stability with negligible

current change was observed up to 25 h. The inset in Fig. 6c shows a corresponding digital photograph of electrolyzer system, where a large number of bubbles were produced on the surface of both electrodes, indicating that as-anodic $CoS_x$ indeed concurrently catalyzed both HER and OER.

## 4. Conclusions

In summary, for the first time, we have prepared a porous cobalt sulfides ($CoS_x$) nanostructure directly on Co foil via a facile, cheap anodization method. The self-grown as-anodic $CoS_x$ demonstrates an outstanding electrochemical performance towards both HER and OER, requiring an overpotential of 102 and 284 mV at a current density of 10 mA m$^{-2}$ in 1.0 M KOH, respectively. Additionally, to achieve a current density of 10 mA m$^{-2}$ for overall water splitting, the as-anodic $CoS_x$||as-anodic $CoS_x$ electrolyzer requires a voltage of 1.64 V, which is comparable to noble Pt||RuO$_2$ (1.61 V) and outperforms most of the previous cobalt sulfide-based bifunctional catalysts. The outstanding electrochemical properties can be assigned to the unique mesoporous architecture, which allows abundant active atoms on the surface for catalytic reactions. Moreover, the robust combination of the $CoS_x$ with the substrate would further facilitate electron transfer across the $CoS_x$/Co interface, as well as maintain its structur after long-time test. Given the simplicity and low-energy/time consumable of the anodization strategy, the work can help us to construct a big family of transition metal chalcogenides ($MS_x$, where M = W, Fe, Mo, Co, Ni, Nb or V) as bifunctional catalysts towards overall water splitting in industry.

**Supporting Information**

Supporting Information is available online or from the author.

**Conflict of Interest**

The authors declare no conflict of interest.

# Acknowledgements

We acknowledge the supports of the National Natural Science Foundation of China (Project No. 21905180), and the China Postdoctoral Science Foundation (No. 2019M663055). This work was jointly supported by the National Natural Science Foundation of China (Project No. 51778369), the Guangdong Provincial Key Laboratory of Energy Materials for Electric Power (Project No. 2018B030322001). We appreciate Instrumental Analysis Center of Shenzhen University (Xili Campus) for providing some equipment for material characterization.

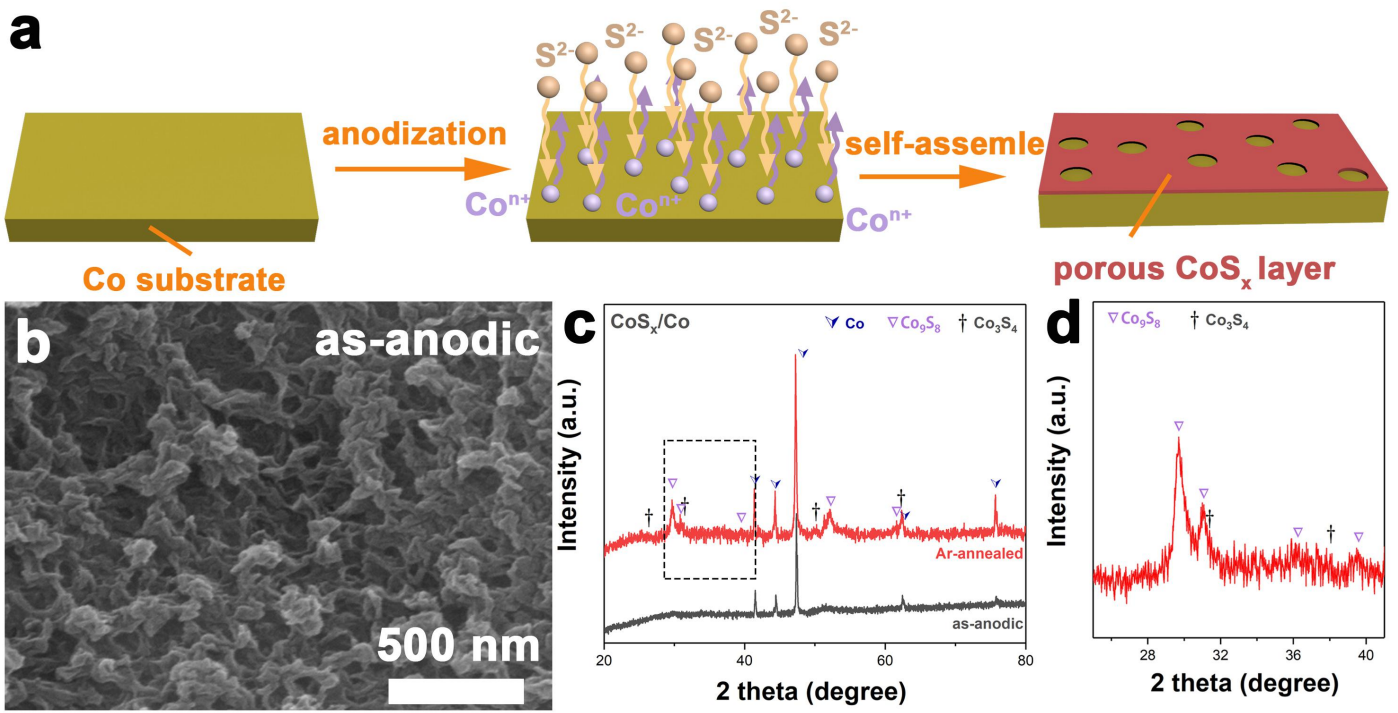

Fig. 1 a) Schematic illustration of the anodization process of $CoS_x$ on the metal substrate. b) SEM image of as-anodic $CoS_x$ obtained *via* anodizing Co foil in 0.05 M $Na_2S$-contained EG solution at 10 V for 5 min. c-d) XRD patterns of as-anodic and Ar-annealed $CoS_x$/Co.

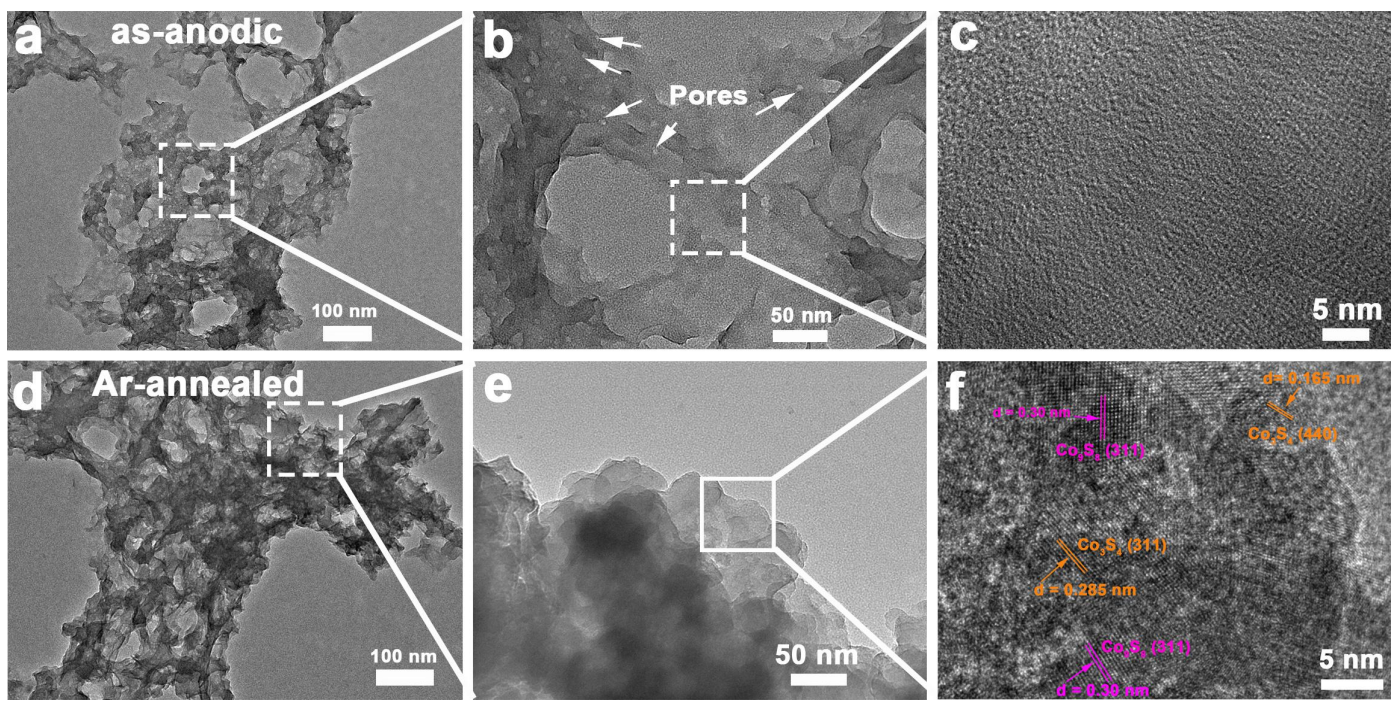

Fig. 2 a-b) TEM and c) HRTEM images of as-anodic CoS$_x$. d-e) TEM and f) HRTEM images of Ar-annealed CoS$_x$. The arrows in (b) indicate the tiny pores.

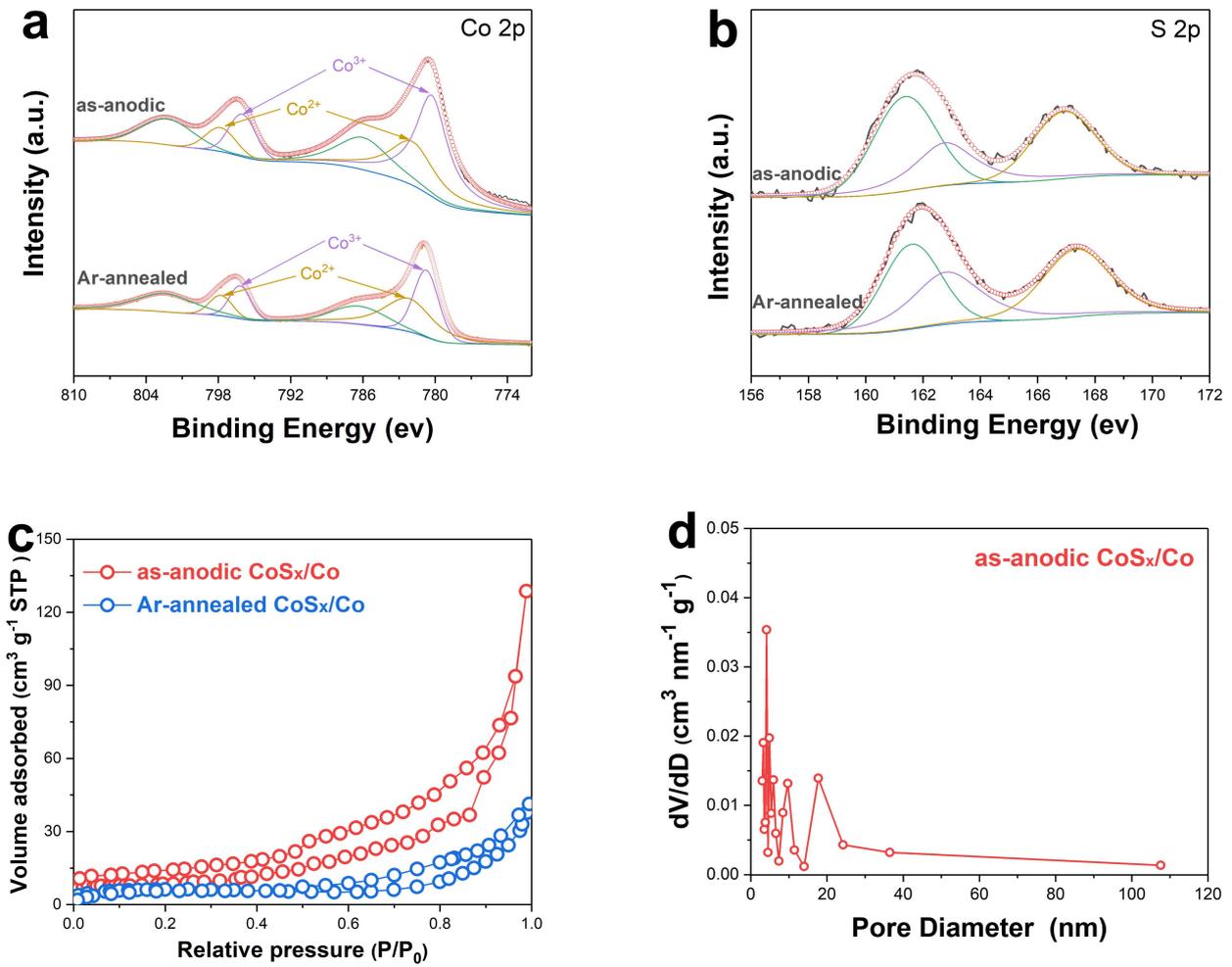

Fig. 3 High-resolution a) Co 2p and b) S 2p XPS spectra of as-anodic and Ar-annealed $CoS_x$. c) Nitrogen adsorption-desorption isotherms of as-anodic and Ar-annealed $CoS_x$. d) Pore size distribution of as-anodic $CoS_x$.

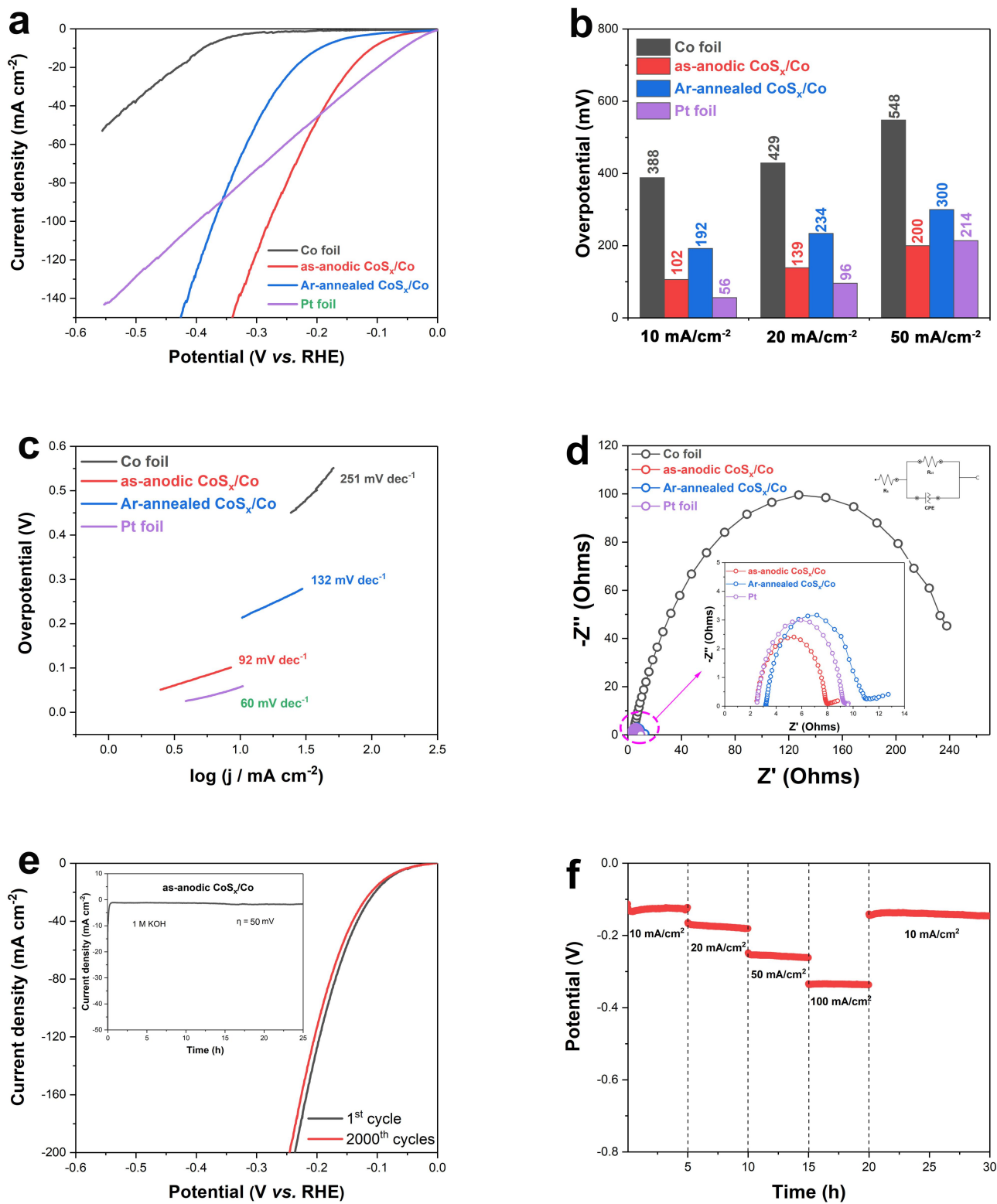

Fig. 4 HER electrocatalytic performance of the prepared catalysts. a) IR-corrected polarization curves of as-anodic $CoS_x$, Ar-annealed $CoS_x$, pure Co and Pt in 1 M KOH at a scan rate of 2 mV/s. b) Obtained overpotentials to achieve different current densities (10, 20 and 50 mA cm$^{-2}$), and c) the corresponding Tafel plots from the polarization curves. d) Nyquist plots of electrochemical impedance spectra of different catalysts at an overpotential of 250 mV for HER process and the corresponding equivalent circuit model (inset). e) Polarization curves of as-anodic $CoS_x$ catalyst before and after 2000 CV cycles at 100 mV/s. The inset is the *j-t* curve of as-anodic $CoS_x$ catalyst towards HER at an overpotential of 50 mV for 25 h. f) Electrochemical stability of as-anodic $CoS_x$ catalyst tested by chronopotentiometry at different current densities.

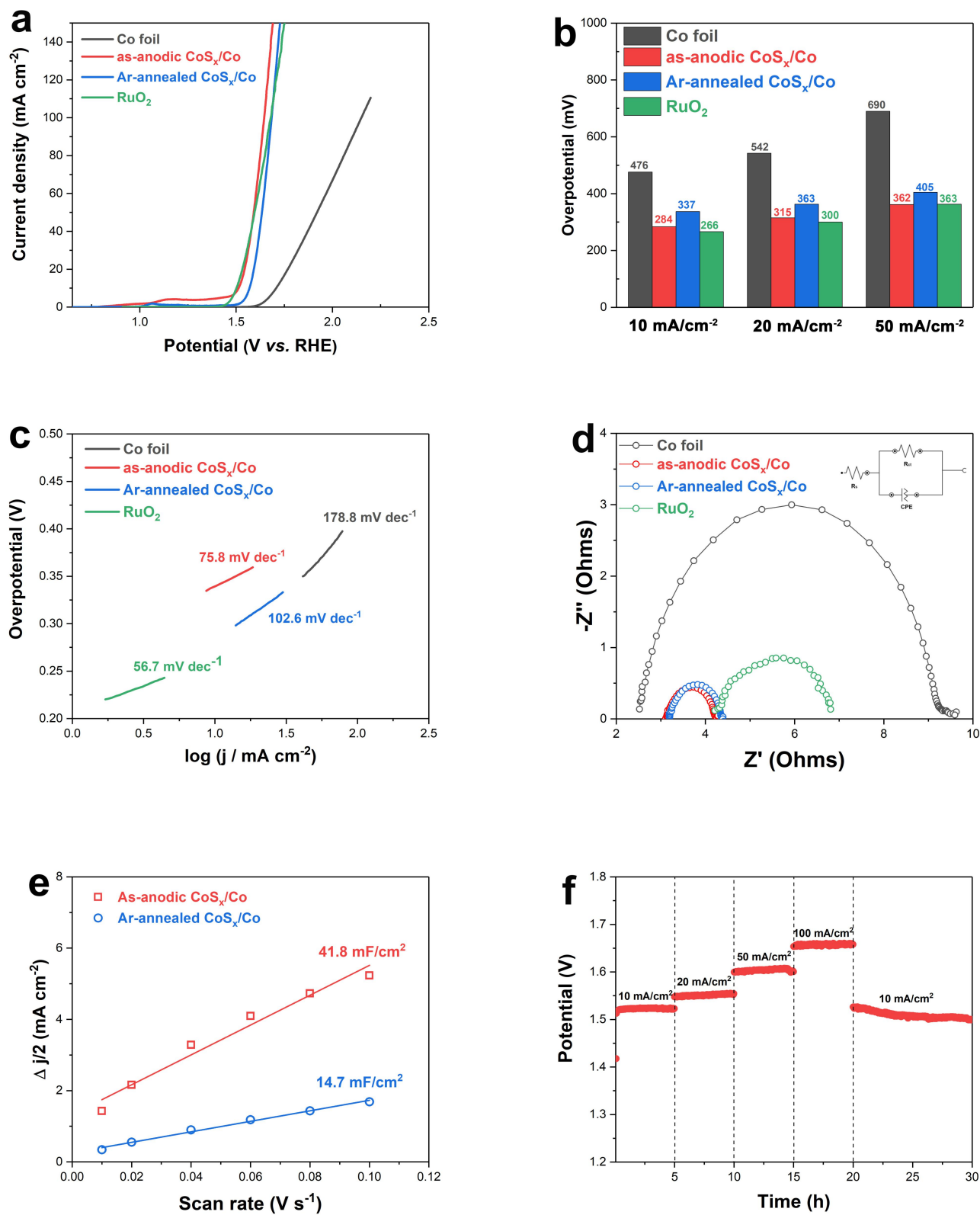

Fig. 5 OER electrocatalytic performance of the prepared catalysts. a) IR-corrected polarization curves of as-anodic $CoS_x$, Ar-annealed $CoS_x$, pure Co and benchmark $RuO_2$ in 1 M KOH at a scan rate of 2 mV/s. b) The overpotentials at different current densities (10, 20 and 50 mA cm$^{-2}$), and c) the corresponding Tafel plots of different catalysts from the polarization curves. d) Nyquist plots of electrochemical impedance spectra of different catalysts at an overpotential of 400 mV for OER process and the corresponding equivalent circuit model (inset). e) Current densities determined at a potential of 0.31 V (vs. RHE) as a function of scan rate for as-anodic and Ar-annealed $CoS_x$ catalysts. f) Electrochemical stability of as-anodic $CoS_x$ catalyst tested by chronopotentiometry at different current densities.

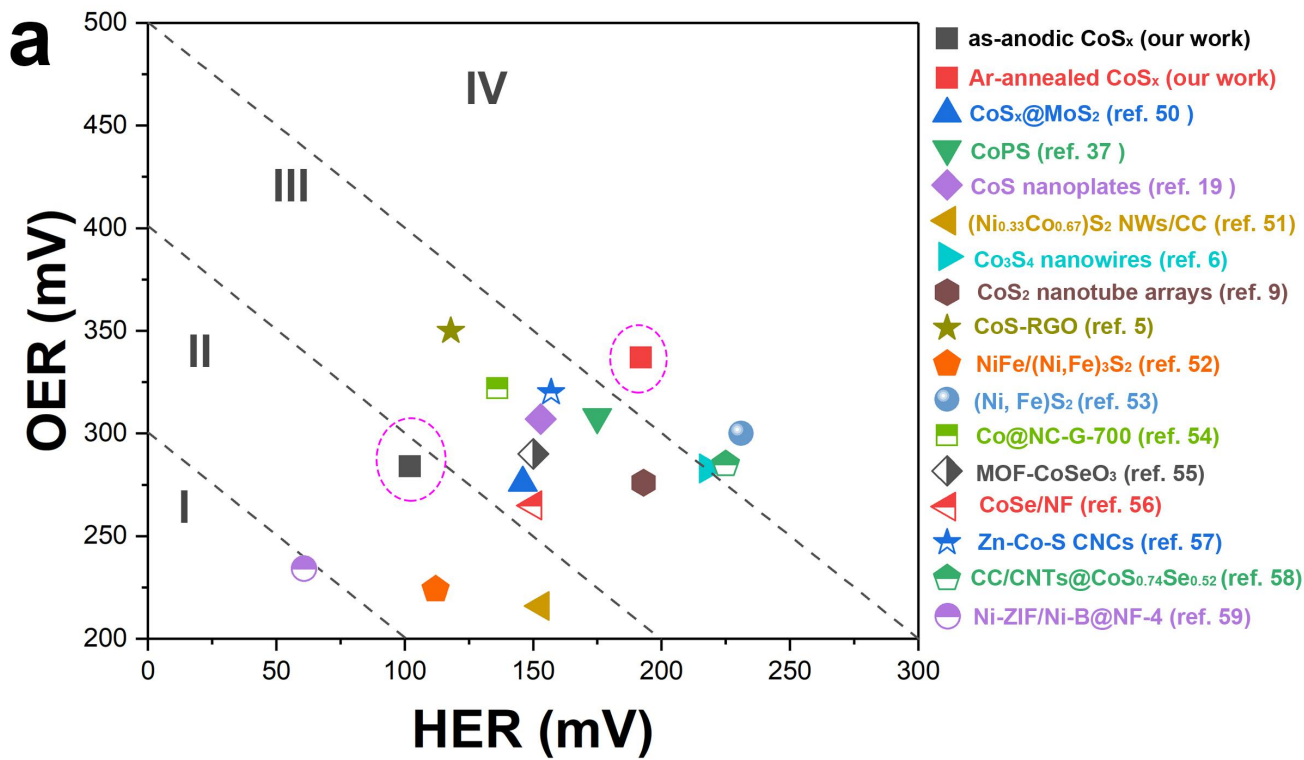

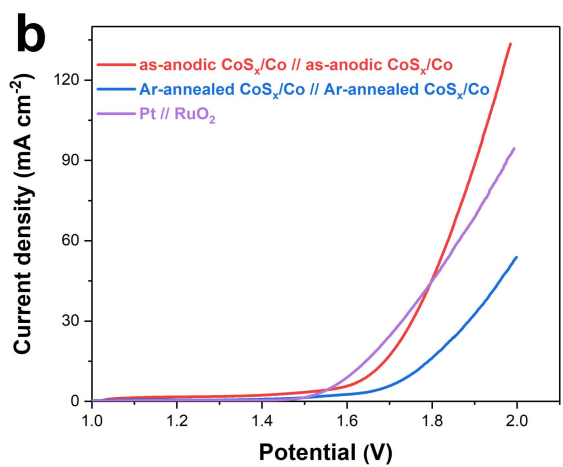
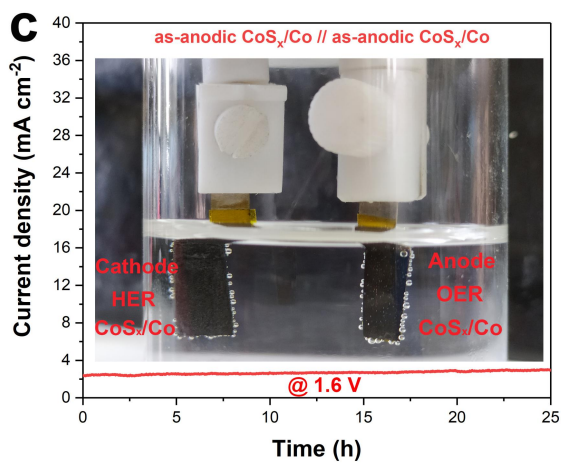

Fig. 6 a) Comparison electrochemical performance ($\eta_{10}$) of some latest reported cobalt-based bifunctional catalysts for HER and OER. b) Polarization curves of as-anodic CoS$_x$‖as-anodic CoS$_x$, Ar-annealed CoS$_x$‖Ar-annealed CoS$_x$, and noble Pt‖RuO$_2$ for overall water splitting in 1 M KOH at a scan rate of 2 mV s$^{-1}$. Prior to polarization curves record, 5 cycles of CV testing at a scan rate of 0.1 V s$^{-1}$ was carried out. c) Long-time duration test of as-anodic CoS$_x$‖as-anodic CoS$_x$ for overall water splitting at a controlled voltage of 1.6 V. The inset was the corresponding photograph of the electrolyzer system.